\documentclass[usenatbib]{aat} 
\begin{document}
\setcounter{page}{1}
\nametom{X(X), \pageref{firstpage}--\pageref{lastpage} (XXXX)}
\newcommand{\adv}{{Adv. Spa. Res.}}
\newcommand{\annG}{{Annales Geophysicae}}
\newcommand{\ag}{{Ann. Geophys.}}
\newcommand{\gafd}{{Geophys. Astrophys. Fluid Dyn.}}
\newcommand{\ijga}{{Int. J. Geomag. Aeron.}}
\newcommand{\jastp}{{J. Atmos. Sol. Terr. Phys.}}
\newcommand{\apj}{ApJ}
\newcommand{\apjs}{ApJS}
\newcommand{\apjl}{ApJL}
\newcommand{\aap}{A{\&}A}
\newcommand{\aaps}{A{\&}AS}
\newcommand{\mnras}{MNRAS}
\newcommand{\aj}{AJ}
\newcommand{\araa}{ARA\&A}
\newcommand{\pasp}{PASP}
\newcommand{\na}{New Astronomy}
\newcommand{\nat}{Nature}
\newcommand{\ssr}{Space Science Reviews}

\title{Looking at Radio-Quiet AGN with Radio Polarimetry}
\author{Silpa S.\inst{1} and P. Kharb\inst{1}}
\institute{National Centre for Radio Astrophysics – Tata Institute of Fundamental Research, S. P. Pune University Campus, Ganeshkhind, Pune 411007, India\\ \email{silpa@ncra.tifr.res.in}}  
\date{Received date}
\titlerunning{Radio-quiet quasars}
\authorrunning{Silpa S. \& P. Kharb} 

\maketitle
\label{firstpage}

\begin{abstract} 
The dominant radio emission mechanism in radio-quiet quasars (RQQs) is an open question. Primary contenders include: low-power radio jets, winds, star-formation and coronal emission. Our work suggests that radio polarization and emission-line studies can help to distinguish between these scenarios and determine the primary contributor. Our multi-frequency, multi-scale radio polarization study has revealed a composite jet and ``wind'' radio outflow in the radio-intermediate quasar, III Zw 2, as well as in the BALQSO, Mrk 231. Our radio polarization study in conjunction with the [O III] emission-line study of five type 2 RQQs have provided insights on the interplay of jets/winds and emission-line gas. These sources reveal an anti-correlation between polarized radio emission and [O III] emission. This is similar to that observed in some radio-loud AGN in the literature and suggests that the radio emission could be depolarized by the emission-line gas. Overall, our work suggests that a close interaction between the radio outflow and the surrounding gaseous environment is likely to be responsible for their stunted form in RQ and RI AGN.
\end{abstract}

\section{Introduction}
The radio-loud/radio-quiet dichotomy has been one of the long-standing problems in AGN physics, which was first identified by \citet{Kellermann89} in the optically-selected Palomar Green (PG) quasar sample \citep{BorosonGreen92}. They observed that the radio-loudness parameter (R), defined by the ratio of the 5~GHz flux density to the optical B-band flux density, followed a bimodal distribution. While only $10-20$\% of the AGN population were found to be radio-loud (RL; R $>$ 10), i.e., those that launch powerful radio jets extending to 100s of kpc or Mpc scales, the vast majority of AGN are radio-quiet (RQ; R $\leq$ 10), hosting small-scale radio jets and/or diffuse and wind-like outflows. The existence of this sharp division has been questioned in the literature. \citet{Falcke96} proposed another class of AGN, namely radio-intermediate (RI) AGN, which had intermediate R values (10 $<$ R $<$ 250) and were identified to be the relativistically boosted counterparts of RQ AGN.

\section{Key questions}
We discuss below some of the key questions pertaining to the RL/RQ division:

(i) What is the origin of the RL/RQ dichotomy?

(ii) What is the origin of radio emission in RQ and RI AGN? 

(iii) What causes stunted radio outflows in RQ and RI AGN? 

There is a general consensus about the origin of radio emission in RL AGN, which is often attributed to the \citet{BlandfordZnajek77} mechanism (hereafter referred to as BZ77) where the relativistic jets are produced by the electromagnetic extraction of rotational energy from spinning black holes (BHs). However, the origin of radio emission in RQ and RI AGN is still a matter of debate. It is not clear if their emission mechanism is that of BZ77 or \citet{BlandfordPayne82} or a combination of both. In the \citet{BlandfordPayne82} mechanism (hereafter referred to as BP82), the angular momentum and energy are magnetically extracted from the accretion disk to power AGN winds. Possible origins of radio emission in RQ AGN as proposed in the literature include low-power jets, winds, star-formation and coronal emission \citep[e.g.,][]{Panessa19}. Recent works have demonstrated the ability of radio polarization studies, in conjunction with spatially resolved emission-line studies, to address fundamental questions related to the nature of radio outflows in AGN \citep{Sebastian19a,Sebastian19b,Sebastian20,Silpa21a,Silpa21b}.

\section{The nature of the radio outflow in III~Zw~2}

\begin{figure*}
\centering
\includegraphics[width=19cm]{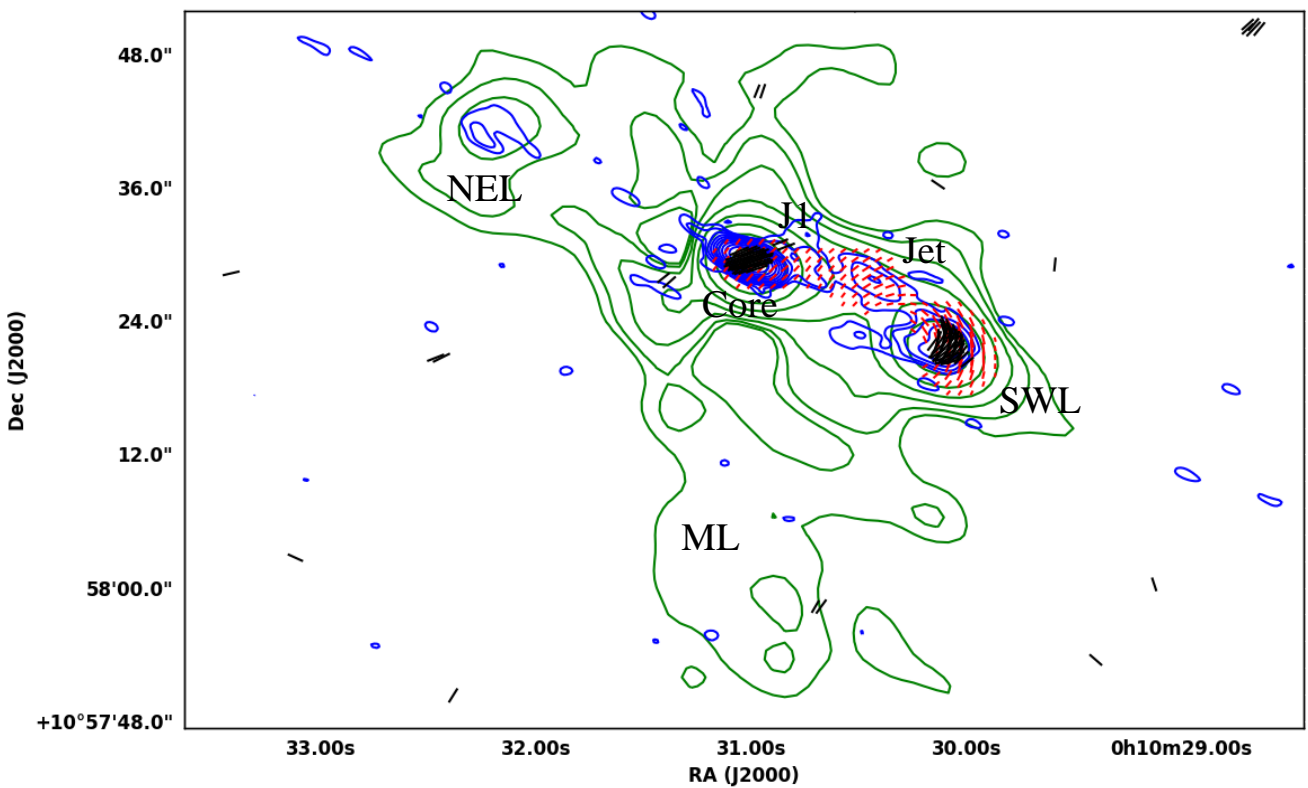}
\caption{The uGMRT 685~MHz total intensity contours in green with inferred B-fields as red ticks and EVLA 5~GHz B-array in blue with inferred B-fields as black ticks, for III~Zw~2 \citep{Silpa21a}. The length of these vectors is proportional to polarized intensity. NEL refers to the north-eastern lobe, SWL refers to the south-western lobe and ML refers to the misaligned lobe in III~Zw~2.}
\label{fig1}
\end{figure*}

III~Zw~2, at a redshift of 0.089, was the first Seyfert galaxy where a superluminal radio jet was discovered \citep{Brunthaler00}. It is also a RI quasar \citep[R=200;][]{Kellermann94} and belongs to the PG quasar sample \citep[PG 0007+106;][]{SchmidtGreen83}. In Figure~\ref{fig1}, we show the uGMRT 685~MHz total intensity contours in green with inferred magnetic (B-) fields as red ticks and VLA 5~GHz B-array total intensity contours in blue with inferred B-fields as black ticks, for III~Zw~2 \citep{Silpa21a}. The polarization electric ($\chi$) vectors have been rotated by 90$\degr$ to obtain the inferred B-fields, assuming optically thin emission \citep{Pacholczyk70}. We note that this assumption is not valid for the optically thick core. However, the vectors have been rotated for illustrative purposes.

We clearly see a triple radio structure comprising of a `hotspot-core-hotspot' in both the uGMRT and VLA images. The curved jet in the VLA image terminates in `bow-shock-like' radio features, instead of compact hotspots typically seen in FRII radio galaxies. The resemblance of this feature with the restarted jet simulations of \citet{ClarkeBurns91} (essentially, their Fig. 1e) suggests restarted jet activity in III Zw 2. The uGMRT image also detects a kpc-scale lobe emission in the south that is misaligned with the primary lobes (annotated as ML in Figure~\ref{fig1}). The misaligned lobe has similar spectral indices and electron lifetimes as the primary lobes, suggesting that it is not a relic lobe. This is reminiscent of the large and `alive' lobes observed in M87 at 330~MHz with the VLA \citep{Owen00}. Moreover, the primary and the misaligned lobes reveal characteristics similar to those seen in the `sputtering' AGN, NGC~3998 \citep{Sridhar20}, such as a mean spectral index between $-0.6$ and $-0.7$ and a lack of clear spectral steepening with distance from the core. This suggests that the AGN is similarly `sputtering' in III Zw 2.

For optically thin regions like jets and lobes, the inferred B-fields are perpendicular to the $\chi$ vectors whereas for optically thick regions like the core, the inferred B-fields are parallel to the $\chi$ vectors \citep{Pacholczyk70}. In the uGMRT image, we find that the inferred B-fields are transverse to the jet direction and aligned with the edges of the south-western lobe. This could either symbolize a series of transverse shocks \citep[e.g.,][]{Gabuzda94,Lister98}, or a toroidal component of a large-scale helical B-field associated with the jet \citep[e.g.,][]{Pushkarev17}. The inferred B-fields in the VLA image are aligned with the lobe edges, as in the uGMRT image, suggesting shock compression. The inferred B-fields in the VLA core are roughly transverse to the local jet direction as well as the VLBA jet direction \citep[since the position angle for the VLBA jet is $\sim$74$\degr$;][]{Brunthaler00,Brunthaler05}. The inferred B-fields in the VLA jet component J1 are parallel to the local jet direction, and hence could represent the poloidal component of the helical jet B-field.

Alternately, the transverse B-fields in the uGMRT image may connote toroidal B-fields threading an AGN wind \citep[e.g.,][]{Miller12,MehdipourConstantini19}, sampled on larger spatial scales than sampled by the VLA observations. This wind component could either be an accretion disk wind \citep[e.g.,][]{BlandfordPayne82} or the outer layers of a broadened jet \citep[like a jet sheath, e.g.,][]{Mukherjee18} or a mixture of both. In keeping with this model, the parallel B-fields in the VLA component J1 may represent poloidal B-fields threading the spine of a jet, since poloidal fields are easier to anchor in thick disks which are essential for launching of jets \citep{Miller12}.

Overall, we find a coalescent radio outflow in III~Zw~2 comprising of a jet with poloidal  B-fields immersed inside a magnetized wind with toroidal B-fields. The transverse B-fields in the core could suggest a toroidal component at the base of this composite outflow. Our findings demonstrate the ability of radio polarimetry to probe different layers of the outflow like the jet spine or the jet sheath/wind sampled at different spatial scales and having characteristic B-field structures.

\begin{figure*}
\centering{
\includegraphics[width=16cm]{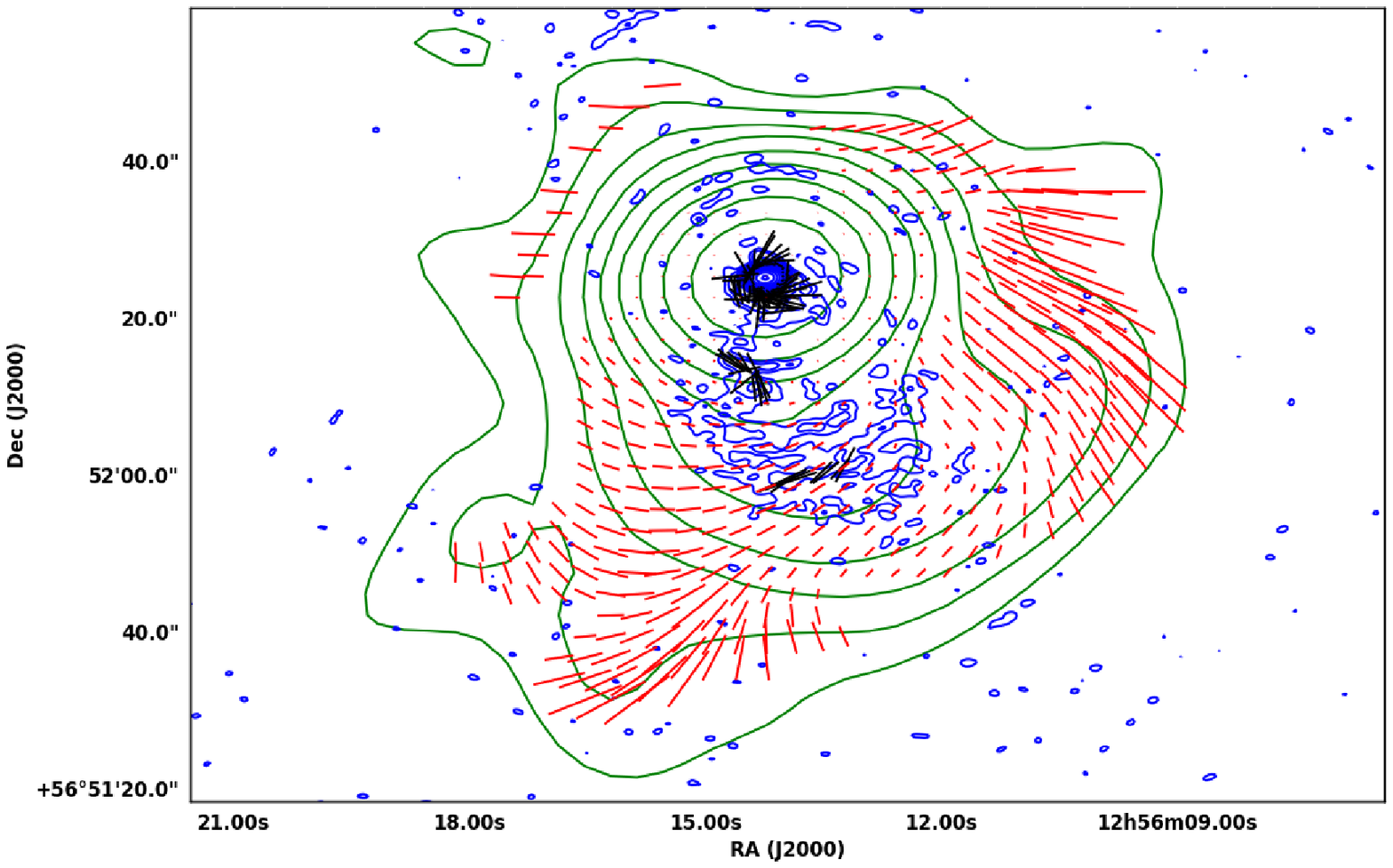}
\includegraphics[width=16cm]{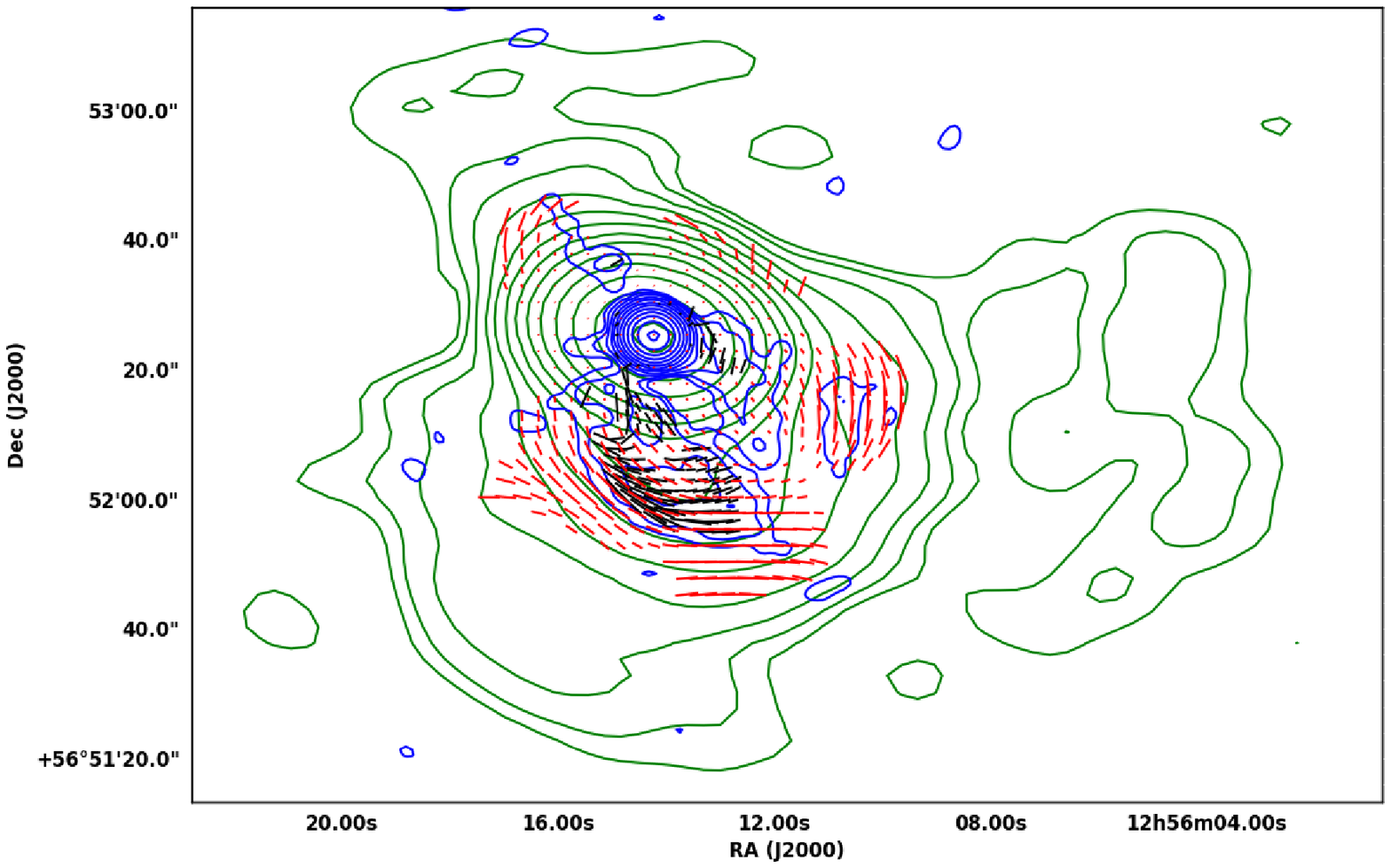}}
\caption{Top: VLA 1.4 GHz total intensity contours for C-array in green with inferred B-fields as red ticks and for A-array in blue with inferred B-fields as black ticks, for Mrk 231. Bottom: VLA 4.9 GHz total intensity contours for D-array in green with inferred B-fields as red ticks and for C-array in blue with inferred B-fields as black ticks, for Mrk 231 \citep{Silpa21b}. The length of these vectors is proportional to the fractional polarization (i.e. ratio of polarized intensity to total intensity). }
\label{fig2}
\end{figure*}

\section{The multi-component outflow in Mrk~231}

Mrk~231, at a redshift of 0.04217, is a well-studied Seyfert type 1 galaxy \citep{Sanders88b,Surace98} and a Broad Absorption Line Quasi-Stellar Object \citep[BALQSO;][]{Boksenberg77,Smith95,Gallagher02}. It is also known to be transitioning from the RQ to RL state \citep{Reynolds17}. The top and bottom panels of Figure~\ref{fig2} show the historical VLA images of Mrk 231 at 1.4 GHz and 4.9 GHz, respectively. Both panels present total intensity contours at lower resolution ($\sim10\arcsec$ scale) in green and higher resolution ($\sim1-4\arcsec$ scale) in blue \citep{Silpa21b}. The vectors in both panels represent inferred B-fields obtained by rotating $\chi$ vectors by 90$\degr$ assuming optically thin emission. Again, this assumption is not valid for the core, but the rotated vectors have been shown for illustrative purposes. 

The $\sim10\arcsec$ scale images detect a diffuse lobe-like emission to the south that extends to $\sim55$ kpc whereas the $\sim1-4\arcsec$ scale images detect a poorly collimated jet extending $\sim25-30$ kpc to the south. The higher resolution images detect a region close to the core with inferred B-fields parallel to the local jet direction. The lower resolution images detect a transverse B-field all the way from the core to the edges of the southern lobe. This is similar to the case of III~Zw~2, in that the poloidal inferred B-fields thread the spine of the weakly collimated jet sampled at higher resolution and toroidal inferred B-fields thread the AGN wind sampled at lower resolution. Thus, Mrk 231 also seems to host a composite jet and wind radio outflow. 

We also find, based on a rotation measure (RM) analysis, that the radio outflow in Mrk 231 becomes increasingly matter-dominated away from the core. Essentially, the magnetization parameter ($\beta$), which is the ratio of gas pressure to magnetic pressure, is found to increase with distance from the core. This is consistent with the mixing of synchrotron plasma with the entrained gas as the radio outflow (jet/lobe and/or wind) propagates through the ambient medium. Thus, one cannot neglect the role of the local environment while studying AGN outflows.

We note that the inferred B-field vectors have not been corrected for Faraday rotation. Given an average RM of about $5-10$~rad~m$^{-2}$ in the southern lobes of III~Zw~2 and Mrk~231, the expected change in the electric vector polarization angle (EVPA) is $<15$~degrees. However, the RM values in certain regions of the lobes are higher, which makes the expected changes in the EVPA also higher. Hence, it is difficult to obtain an unambiguous interpretation of the B-field direction w.r.t to the local radio outflow direction in these sources.

\section{An interplay of jet/wind \& emission-line gas}

In a study carried out on a small sample of type 2 RQ quasars with EVLA polarization and HST [O III] emission-line imaging, we have found a correlation between radio emission (from jets or winds) and [O III] emission but an anti-correlation between radio polarized emission and [O III] emission in these sources. This is similar to that observed in some RL AGN in the literature \citep{vanBreugel84,Heckman84,vanBreugel85} and suggests that the radio emission is depolarized by the emission-line gas. Furthermore, we have found evidence for the mixing of entrained thermal gas and synchrotron plasma in the lobes of these sources. This results in the internal depolarization of the lobe emission in them (Silpa et al. 2022, MNRAS, submitted). Thus, both external and internal depolarization seem to be operational in these RQ quasars. 

\begin{figure*}
\centering
\includegraphics[width=16cm]{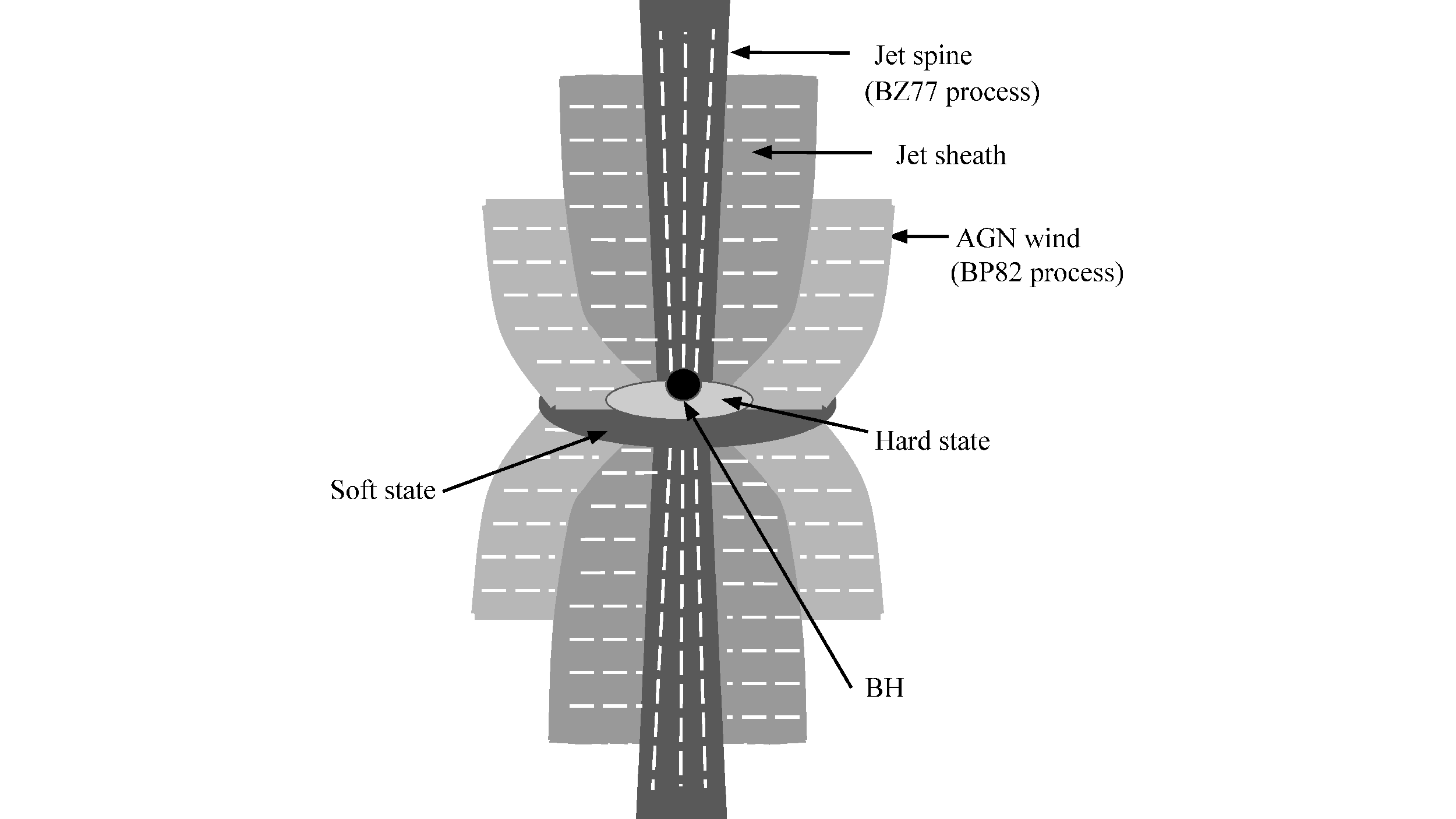} 
\caption{A cartoon to illustrate the relation between the possible changing spectral states of the accretion disk and the multi-component outflow in AGN, as inferred from our work on RQ and RI AGN \citep{Silpa21a}. The white ticks represent the B-field vectors.}
\label{fig3}
\end{figure*}

\section{Implications for the RL-RQ dichotomy}

Overall, two important conclusions can be drawn from our work: (i) a spatial stratification of B-fields is observed in our sources, which signify the presence of more than one mode of radio outflow, (ii) the interaction between various outflow components, as well as the surrounding gaseous environment, is likely to be responsible for the stunted outflows in RQ and RI AGN. We present a cartoon model in Figure~\ref{fig3} to summarize a connection between possible spectral state changes of the accretion disk and the multi-component outflow in AGN \citep{Silpa21a}.

Drawing analogies from X-ray binaries, we propose that the outer regions of the accretion disk in soft state, characterized by the \citet{ShakuraSunyaev73} disk, launches wind (either driven by the BP82 mechanism or it could be a jet sheath), and the inner regions of the accretion disk in hard state, characterized by an advection-dominated accretion flow \citep{NarayanYi94}, launches jet (or a jet spine, driven by the BZ77 mechanism). Such a co-axial outflow would appear co-spatial in projection, as seen in our sources. The interaction of the jet with the wind plasma (possibly occurring in the intervening region between BZ77- and BP82-driven components) may disrupt the jet. Alternately, most of the mass and energy may be carried away by the wind, rather than by the jet. This could explain why the jets are low-powered and small-scaled in RI AGN, as compared to those in RL AGN.

In regular RL and RQ AGN, the transition between the hard and soft spectral states typically occurs on time scales of $10^6-10^8$ yr \citep{AlexanderLeahy87,EnsslinGopal-Krishna01,Shabala08}. However, this may occur on much shorter time scales, such as decades, in sources showing intermittent/sputtering AGN activity. Such rapid transitions may explain why one finds a combination of a jet with moderate radio luminosity and an accretion disk wind in such sources. Furthermore, this model favours the idea of radio-loudness being a function of the epoch at which a source is observed \citep[e.g.,][]{Nipoti05,K-B20,Nyland20}. Its implications are that the RQ AGNs are the ones in their soft state, hosting strong winds and suppressed jets, and RL AGNs are the ones in their hard state, launching powerful jets, at the time of the observations.

\section{Acknowledgements}
We thank the staff of the GMRT who made these observations possible. GMRT is run by the National Centre for Radio Astrophysics of the Tata Institute of Fundamental Research. We acknowledge the support of the Department of Atomic Energy, Government of India, under the project 12-R\&D-TFR-5.02-0700. The National Radio Astronomy Observatory is a facility of the National Science Foundation operated under cooperative agreement by Associated Universities, Inc.


%
%
%
%

\bibliographystyle{aat}
\bibliography{aat}

\begin{thebibliography}{42}
\providecommand{\natexlab}[1]{#1}
\providecommand*{\BibUrl}[1]{\url{#1}}
\expandafter\ifx\csname urlstyle\endcsname\relax
  \providecommand{\doi}[1]{doi:\discretionary{}{}{}#1}\else
  \providecommand{\doi}{doi:\discretionary{}{}{}\begingroup
  \urlstyle{rm}\Url}\fi
\providecommand{\eprint}[2][]{\url{#2}}

\bibitem[{{Alexander} and {Leahy}(1987)}]{AlexanderLeahy87}
{Alexander} P., {Leahy} J.P., 1987.
\newblock \mnras, vol.~ 225, pp.~ 1--26.
\doi{10.1093/mnras/225.1.1}.

\bibitem[{{Blandford} and {Payne}(1982)}]{BlandfordPayne82}
{Blandford} R.D., {Payne} D.G., 1982.
\newblock \mnras, vol.~ 199, pp.~ 883--903.
\doi{10.1093/mnras/199.4.883}.

\bibitem[{{Blandford} and {Znajek}(1977)}]{BlandfordZnajek77}
{Blandford} R.D., {Znajek} R.L., 1977.
\newblock \mnras, vol.~ 179, pp.~ 433--456.
\doi{10.1093/mnras/179.3.433}.

\bibitem[{{Boksenberg} et~al.(1977){Boksenberg}, {Carswell}, {Allen}
  et~al.}]{Boksenberg77}
{Boksenberg} A., {Carswell} R.F., {Allen} D.A., et~al., 1977.
\newblock \mnras, vol.~ 178, pp.~ 451--466.
\doi{10.1093/mnras/178.3.451}.

\bibitem[{{Boroson} and {Green}(1992)}]{BorosonGreen92}
{Boroson} T.A., {Green} R.F., 1992.
\newblock \apjs, vol.~~80, p.~ 109.
\doi{10.1086/191661}.

\bibitem[{{Brunthaler} et~al.(2000){Brunthaler}, {Falcke}, {Bower}
  et~al.}]{Brunthaler00}
{Brunthaler} A., {Falcke} H., {Bower} G.C., et~al., 2000.
\newblock \aap, vol.~ 357, pp.~ L45--L48 (\eprint {arXiv} {astro-ph/0004256}).

\bibitem[{{Brunthaler} et~al.(2005){Brunthaler}, {Falcke}, {Bower}
  et~al.}]{Brunthaler05}
{Brunthaler} A., {Falcke} H., {Bower} G.C., et~al., 2005.
\newblock \aap, vol.~ 435, no.~~2, pp.~ 497--506.
\doi{10.1051/0004-6361:20042427}.

\bibitem[{{Clarke} and {Burns}(1991)}]{ClarkeBurns91}
{Clarke} D.A., {Burns} J.O., 1991.
\newblock \apj, vol.~ 369, p.~ 308.
\doi{10.1086/169762}.

\bibitem[{{En{\ss}lin} and {Gopal-Krishna}(2001)}]{EnsslinGopal-Krishna01}
{En{\ss}lin} T.A., {Gopal-Krishna}, 2001.
\newblock \aap, vol.~ 366, pp.~ 26--34.
\doi{10.1051/0004-6361:20000198} (\eprint {arXiv} {astro-ph/0011123}).

\bibitem[{{Falcke} et~al.(1996){Falcke}, {Patnaik} and {Sherwood}}]{Falcke96}
{Falcke} H., {Patnaik} A.R., {Sherwood} W., 1996.
\newblock \apjl, vol.~ 473, p.~ L13.
\doi{10.1086/310386} (\eprint {arXiv} {astro-ph/9610049}).

\bibitem[{{Gabuzda} et~al.(1994){Gabuzda}, {Mullan}, {Cawthorne}, {Wardle} and
  {Roberts}}]{Gabuzda94}
{Gabuzda} D.C., {Mullan} C.M., {Cawthorne} T.V., {Wardle} J.F.C., {Roberts}
  D.H., 1994.
\newblock \apj, vol.~ 435, p.~ 140.
\doi{10.1086/174801}.

\bibitem[{{Gallagher} et~al.(2002){Gallagher}, {Brandt}, {Chartas}, {Garmire}
  and {Sambruna}}]{Gallagher02}
{Gallagher} S.C., {Brandt} W.N., {Chartas} G., {Garmire} G.P., {Sambruna} R.M.,
  2002.
\newblock \apj, vol.~ 569, no.~~2, pp.~ 655--670.
\doi{10.1086/339171} (\eprint {arXiv} {astro-ph/0112257}).

\bibitem[{{Heckman} et~al.(1984){Heckman}, {van Breugel} and
  {Miley}}]{Heckman84}
{Heckman} T.M., {van Breugel} W.J.M., {Miley} G.K., 1984.
\newblock \apj, vol.~ 286, pp.~ 509--516.
\doi{10.1086/162626}.

\bibitem[{{Kellermann} et~al.(1994){Kellermann}, {Sramek}, {Schmidt}, {Green}
  and {Shaffer}}]{Kellermann94}
{Kellermann} K.I., {Sramek} R.A., {Schmidt} M., {Green} R.F., {Shaffer} D.B.,
  1994.
\newblock \aj, vol.~ 108, p.~ 1163.
\doi{10.1086/117145}.

\bibitem[{{Kellermann} et~al.(1989){Kellermann}, {Sramek}, {Schmidt}, {Shaffer}
  and {Green}}]{Kellermann89}
{Kellermann} K.I., {Sramek} R., {Schmidt} M., {Shaffer} D.B., {Green} R., 1989.
\newblock \aj, vol.~~98, p.~ 1195.
\doi{10.1086/115207}.

\bibitem[{{Kunert-Bajraszewska} et~al.(2020){Kunert-Bajraszewska},
  {Wo{\l}owska}, {Mooley}, {Kharb} and {Hallinan}}]{K-B20}
{Kunert-Bajraszewska} M., {Wo{\l}owska} A., {Mooley} K., {Kharb} P., {Hallinan}
  G., 2020.
\newblock \apj, vol.~ 897, no.~~2, 128.
\doi{10.3847/1538-4357/ab9598} (\eprint {arXiv} {2007.01590}).

\bibitem[{{Lister} et~al.(1998){Lister}, {Marscher} and {Gear}}]{Lister98}
{Lister} M.L., {Marscher} A.P., {Gear} W.K., 1998.
\newblock \apj, vol.~ 504, no.~~2, pp.~ 702--719.
\doi{10.1086/306112}.

\bibitem[{{Mehdipour} and {Costantini}(2019)}]{MehdipourConstantini19}
{Mehdipour} M., {Costantini} E., 2019.
\newblock \aap, vol.~ 625, A25.
\doi{10.1051/0004-6361/201935205} (\eprint {arXiv} {1903.11605}).

\bibitem[{{Miller} et~al.(2012){Miller}, {Raymond}, {Fabian} et~al.}]{Miller12}
{Miller} J.M., {Raymond} J., {Fabian} A.C., et~al., 2012.
\newblock \apjl, vol.~ 759, no.~~1, L6.
\doi{10.1088/2041-8205/759/1/L6} (\eprint {arXiv} {1208.4514}).

\bibitem[{{Mukherjee} et~al.(2018){Mukherjee}, {Bicknell}, {Wagner},
  {Sutherland} and {Silk}}]{Mukherjee18}
{Mukherjee} D., {Bicknell} G.V., {Wagner} A.Y., {Sutherland} R.S., {Silk} J.,
  2018.
\newblock \mnras, vol.~ 479, no.~~4, pp.~ 5544--5566.
\doi{10.1093/mnras/sty1776} (\eprint {arXiv} {1803.08305}).

\bibitem[{{Narayan} and {Yi}(1994)}]{NarayanYi94}
{Narayan} R., {Yi} I., 1994.
\newblock \apjl, vol.~ 428, p.~ L13.
\doi{10.1086/187381} (\eprint {arXiv} {astro-ph/9403052}).

\bibitem[{{Nipoti} et~al.(2005){Nipoti}, {Blundell} and {Binney}}]{Nipoti05}
{Nipoti} C., {Blundell} K.M., {Binney} J., 2005.
\newblock \mnras, vol.~ 361, no.~~2, pp.~ 633--637.
\doi{10.1111/j.1365-2966.2005.09194.x} (\eprint {arXiv} {astro-ph/0505280}).

\bibitem[{{Nyland} et~al.(2020){Nyland}, {Dong}, {Patil} et~al.}]{Nyland20}
{Nyland} K., {Dong} D.Z., {Patil} P., et~al., 2020.
\newblock arXiv e-prints, arXiv:2011.08872 (\eprint {arXiv} {2011.08872}).

\bibitem[{{Owen} et~al.(2000){Owen}, {Eilek} and {Kassim}}]{Owen00}
{Owen} F.N., {Eilek} J.A., {Kassim} N.E., 2000.
\newblock \apj, vol.~ 543, no.~~2, pp.~ 611--619.
\doi{10.1086/317151} (\eprint {arXiv} {astro-ph/0006150}).

\bibitem[{{Pacholczyk}(1970)}]{Pacholczyk70}
{Pacholczyk} A.G., 1970.
\newblock {Radio astrophysics. Nonthermal processes in galactic and
  extragalactic sources}.

\bibitem[{{Panessa} et~al.(2019){Panessa}, {Baldi}, {Laor} et~al.}]{Panessa19}
{Panessa} F., {Baldi} R.D., {Laor} A., et~al., 2019.
\newblock Nature Astronomy, vol.~~3, pp.~ 387--396.
\doi{10.1038/s41550-019-0765-4} (\eprint {arXiv} {1902.05917}).

\bibitem[{{Pushkarev} et~al.(2017){Pushkarev}, {Kovalev}, {Lister}
  et~al.}]{Pushkarev17}
{Pushkarev} A., {Kovalev} Y., {Lister} M., et~al., 2017.
\newblock Galaxies, vol.~~5, no.~~4, p.~~93.
\doi{10.3390/galaxies5040093} (\eprint {arXiv} {1712.03025}).

\bibitem[{{Reynolds} et~al.(2017){Reynolds}, {Punsly}, {Miniutti}, {O'Dea} and
  {Hurley-Walker}}]{Reynolds17}
{Reynolds} C., {Punsly} B., {Miniutti} G., {O'Dea} C.P., {Hurley-Walker} N.,
  2017.
\newblock \apj, vol.~ 836, no.~~2, 155.
\doi{10.3847/1538-4357/836/2/155} (\eprint {arXiv} {1701.03190}).

\bibitem[{{Sanders} et~al.(1988){Sanders}, {Soifer}, {Elias}, {Neugebauer} and
  {Matthews}}]{Sanders88b}
{Sanders} D.B., {Soifer} B.T., {Elias} J.H., {Neugebauer} G., {Matthews} K.,
  1988.
\newblock \apjl, vol.~ 328, p.~ L35.
\doi{10.1086/185155}.

\bibitem[{{Schmidt} and {Green}(1983)}]{SchmidtGreen83}
{Schmidt} M., {Green} R.F., 1983.
\newblock \apj, vol.~ 269, pp.~ 352--374.
\doi{10.1086/161048}.

\bibitem[{{Sebastian} et~al.(2019{\natexlab{a}}){Sebastian}, {Kharb}, {O'Dea},
  {Colbert} and {Baum}}]{Sebastian19b}
{Sebastian} B., {Kharb} P., {O'Dea} C.P., {Colbert} E.J.M., {Baum} S.A.,
  2019{\natexlab{a}}.
\newblock \apj, vol.~ 883, no.~~2, 189.
\doi{10.3847/1538-4357/ab371a} (\eprint {arXiv} {1907.12765}).

\bibitem[{{Sebastian} et~al.(2019{\natexlab{b}}){Sebastian}, {Kharb}, {O'Dea},
  {Gallimore} and {Baum}}]{Sebastian19a}
{Sebastian} B., {Kharb} P., {O'Dea} C.P., {Gallimore} J.F., {Baum} S.A.,
  2019{\natexlab{b}}.
\newblock \mnras, vol.~ 490, no.~~1, pp.~ L26--L31.
\doi{10.1093/mnrasl/slz136} (\eprint {arXiv} {1908.09396}).

\bibitem[{{Sebastian} et~al.(2020){Sebastian}, {Kharb}, {O'Dea}, {Gallimore}
  and {Baum}}]{Sebastian20}
{Sebastian} B., {Kharb} P., {O'Dea} C.P., {Gallimore} J.F., {Baum} S.A., 2020.
\newblock \mnras, vol.~ 499, no.~~1, pp.~ 334--354.
\doi{10.1093/mnras/staa2473} (\eprint {arXiv} {2008.06039}).

\bibitem[{{Shabala} et~al.(2008){Shabala}, {Ash}, {Alexander} and
  {Riley}}]{Shabala08}
{Shabala} S.S., {Ash} S., {Alexander} P., {Riley} J.M., 2008.
\newblock \mnras, vol.~ 388, no.~~2, pp.~ 625--637.
\doi{10.1111/j.1365-2966.2008.13459.x} (\eprint {arXiv} {0805.4152}).

\bibitem[{{Shakura} and {Sunyaev}(1973)}]{ShakuraSunyaev73}
{Shakura} N.I., {Sunyaev} R.A., 1973.
\newblock \aap, vol.~ 500, pp.~ 33--51.

\bibitem[{{Silpa} et~al.(2021{\natexlab{a}}){Silpa}, {Kharb}, {Harrison}
  et~al.}]{Silpa21a}
{Silpa} S., {Kharb} P., {Harrison} C.M., et~al., 2021{\natexlab{a}}.
\newblock \mnras, vol.~ 507, no.~~1, pp.~ 991--1001.
\doi{10.1093/mnras/stab1870} (\eprint {arXiv} {2107.01818}).

\bibitem[{{Silpa} et~al.(2021{\natexlab{b}}){Silpa}, {Kharb}, {O'Dea}
  et~al.}]{Silpa21b}
{Silpa} S., {Kharb} P., {O'Dea} C.P., et~al., 2021{\natexlab{b}}.
\newblock \mnras, vol.~ 507, no.~~2, pp.~ 2550--2561.
\doi{10.1093/mnras/stab2110} (\eprint {arXiv} {2107.09466}).

\bibitem[{{Smith} et~al.(1995){Smith}, {Schmidt}, {Allen} and
  {Angel}}]{Smith95}
{Smith} P.S., {Schmidt} G.D., {Allen} R.G., {Angel} J.R.P., 1995.
\newblock \apj, vol.~ 444, p.~ 146.
\doi{10.1086/175589}.

\bibitem[{{Sridhar} et~al.(2020){Sridhar}, {Morganti}, {Nyland}
  et~al.}]{Sridhar20}
{Sridhar} S.S., {Morganti} R., {Nyland} K., et~al., 2020.
\newblock \aap, vol.~ 634, A108.
\doi{10.1051/0004-6361/201936796} (\eprint {arXiv} {1912.04812}).

\bibitem[{{Surace} et~al.(1998){Surace}, {Sanders}, {Vacca}, {Veilleux} and
  {Mazzarella}}]{Surace98}
{Surace} J.A., {Sanders} D.B., {Vacca} W.D., {Veilleux} S., {Mazzarella} J.M.,
  1998.
\newblock \apj, vol.~ 492, no.~~1, pp.~ 116--136.
\doi{10.1086/305028}.

\bibitem[{{van Breugel} et~al.(1984){van Breugel}, {Heckman} and
  {Miley}}]{vanBreugel84}
{van Breugel} W., {Heckman} T., {Miley} G., 1984.
\newblock \apj, vol.~ 276, pp.~ 79--91.
\doi{10.1086/161594}.

\bibitem[{{van Breugel} et~al.(1985){van Breugel}, {Miley}, {Heckman},
  {Butcher} and {Bridle}}]{vanBreugel85}
{van Breugel} W., {Miley} G., {Heckman} T., {Butcher} H., {Bridle} A., 1985.
\newblock \apj, vol.~ 290, pp.~ 496--516.
\doi{10.1086/163007}.

\end{thebibliography}

\label{lastpage}
\end{document}